\documentclass[useAMS,usenatbib]{mn2e}

\usepackage{graphicx}

\title[Astroclimatology at Paranal and ORM]{The astroclimatological comparison of the Paranal Observatory and El Roque de Los Muchachos Observatory}
\author[G. Lombardi et al.]{G. Lombardi$^{1}$$,$$^{2}$$,$$^{3}$\thanks{E-mail:
glombard@eso.org}, V. Zitelli$^{2}$ and S. Ortolani$^{4}$\\
$^{1}$European Southern Observatory, Av. Alonso de Cordova 3107, Santiago 19, Chile\\
$^{2}$National Institute for Astrophysics, Bologna Astronomical Observatory, via Ranzani 1, I-40127, Bologna, Italy\\
$^{3}$Department of Astronomy, University of Bologna, via Ranzani 1, I-40127, Bologna, Italy\\
$^{4}$Department of Astronomy, University of Padova, Vicolo dell'Osservatorio 3, I-35122, Padova, Italy}

\begin{document}

\date{Accepted 2009 June 25. Received 2009 June 15; in original form 2009 January 11}

\pagerange{\pageref{firstpage}--\pageref{lastpage}} \pubyear{2009}

\maketitle

\label{firstpage}

\begin{abstract}
The new extremely large telescope projects need accurate evaluation of the candidate sites. 
In this paper we present the astroclimatological comparison
  between the Paranal Observatory, located on the coast of the Atacama Desert (Chile), and the Observatorio del Roque de Los Muchachos (ORM),
  located in La Palma (Canary Islands). We apply a statistical analysis using long term databases from Paranal and Carlsberg Meridian Telescope
  (CAMC) weather stations.
  The monthly, seasonal and annual averages of the main synoptical parameters in the two sites are computed. We compare the long term trends
  in order to understand the main differences between the two sites. Significant differences between the two analyzed sites have been found.
  Temperature have increasing trends in
  both observatories with somewhat higher evidence at ORM. Seasonal variations of pressure at Paranal have been
  highly decreasing since 1989 and we do not see the same phenomenon at ORM. The two sites are dominated by high pressure.
  In cold seasons $RH$ is lower than 60\% at CAMC and $15\%$ at Paranal. In warm seasons $RH$ is lower than 40\% at CAMC and $20\%$ at Paranal.
  The analysis of the dew point has shown better conditions at Paranal with respect to CAMC in winter,
  autumn and spring before 2001, while the two sites are becoming similar afterwards. Winds at ORM are subject to pronunced local variations.
\end{abstract}


\section{Introduction}
This study follows a series of papers aimed to compare two of the most
important sites candidates to host the future extremely large ground
based telescopes. The analysis of the astronomical and
meteorological parameters in different sites are conducted, since
today, using different approach.
%Thanks to the long series of synoptical data collected up to now
%in each interested site, it is possible to have more
%homogeneous and reliable comparison of these site  using  the
%same statistical approach; in particular
In this work we want to apply same methods to compare two different and important sites
for the development of the astronomical observations: the site of
Paranal Observatory located in the Atacama Desert (Chile, southern hemisphere),
and the Observatorio del Roque de Los
Muchachos (ORM) located at La Palma (Canary Islands, northern
hemisphere). The two opposite locations have different observing
characteristics not only because it is possible to see two different
 sky regions,  but also because the oceanic currents impress different
climatic regimes to the two sites. While Paranal is located in a
desert area, 12 km far from the Pacific Ocean, and it is subjected
to the oscillating weather variation due the presence of El Ni\~no
and La Ni\~na events (Sarazin\citealt{sarazin97}), the ORM is influenced by a semipermanent
Azores high pressure system and it is influenced by the almost
periodical variation of the North Atlantic Oscillation (NAO) (Wanner et al.\citealt{wanner01}; Graham\citealt{graham05}; Lombardi et al.\citealt{lombardi06}).\\
%northern Chile, at an altitude of 2636 m a.s.l., and the
%Observatorio del Roque de Los Muchachos (ORM) located at La Palma
%(Canary Islands) at $\sim$ $2350$ m a.s.l..
%Both the two sites have collected a long series of synoptic data.
More than 20 years of meteorological data have been collected at ORM
using the Carlsberg Automatic Meridian Circle (CAMC) meteorological
station and a detailed analysis can be found in Lombardi et al.
\cite{lombardi06} (hereafter Paper I) and Lombardi et al.
\cite{lombardi07} (hereafter Paper II). Paper I presents a
complete analysis of the vertical temperature gradients at Telescopio
Nazionale Galileo (TNG) and their
correlation with the astronomical seeing. Instead, Paper II reports
an analysis of the correlation between wind and astronomical
parameters as well as the overall long term weather conditions at
ORM. Differences in the ORM microclimate have been demonstrated in a
detailed comparison between synoptical parameters taken at three
different locations in the observatory on a 1000 m spatial scale.
ORM shows to be dominated by high pressure, and characterized by an
averaged relative humidity lower than 50\%.
Finally, in Lombardi et al. \cite{lombardi08a} we have analysed
in detail the properties of the dust concentration on La Palma from ground based
measurements and estimated the aerosol extinction in $B$, $V$ and $I$ on the site.\\
Extensive site testing campaigns have been  conducted on the top of
the Paranal Observatory since years. Thanks to the excellent
results, the site was chosen to host the four Very Large Telescopes
(VLT) by the European Southern Observatory (ESO).\\
This site, like La Silla, the other Chilean ESO site, have been very
deeply analyzed by ESO teams. Now ESO telescopes are considered the
touchstones and their characteristics are used to be compared with the other sites
and the other telescopes.\\
The present paper is organized as follows:
\begin{itemize}
   \item \textbf{Section 2}: describes the Paranal Astronomical Site Monitor and the data reduction;
   \item \textbf{Section 3}: compares the temperature, pressure, relative humidity and
   correlate each parameter with the Southern Oscillation Index (at Paranal) and NAO (at ORM);
   \item \textbf{Section 4}: shows the dew point comparison at the two sites;
   \item \textbf{Section 5}: analysis of wind direction and wind speed;
   %\item \textbf{Section 6}: analysis of the seeing vs. vertical temperature gradient, wind
   %direction and wind speed at Paranal;
   \item \textbf{Section 6}: summarization of the final conclusions.
\end{itemize}
  \begin{table}%[t]
    \begin{center}%\scriptsize
      \caption[]{Geographical positions of Paranal and CAMC.}
   \label{telescopes}\scriptsize
        \begin{tabular}{l | c c c}
          \hline
           & Latitude & Longitude & Height [m a.s.l.]\\
           \hline
 Paranal & 24$^{\circ}$ 37' 31'' S & 70$^{\circ}$ 24' 10'' W & 2636$^{\mathrm{\left[a\right]}}$\\
CAMC & 28$^{\circ}$ 45' 36'' N & 17$^{\circ}$ 52' 57'' W & 2326$^{\mathrm{\left[b\right]}}$\\
           \hline
        \end{tabular}
        \end{center}
        \begin{list}{}{}
        \item{$^{\mathrm{\left[a\right]}}$ Platform altitude.}
        \item{$^{\mathrm{\left[b\right]}}$ Dome floor.}
        \end{list}
  \end{table}
\begin{table}
      \caption{Available databases for the observatories. The data coverage with respect to the Total is also reported.}
    \begin{center}\scriptsize
   \label{databases}
        \begin{tabular}{l | c | c | c }
           \hline
     & TNG & CAMC & Paranal\\
           \hline
   Data rate  & 30 sec & 5 min & 20 min\\
           \hline
   Begin  & March 1998 & May 1984 & January 1985\\
   End  & December 2007 & March 2005 & December 2006\\
   Total  & $\sim$$10$ yr & $\sim$$21$ yr & $\sim$$22$ yr\\
	\hline
   Data coverage  &  &  & \\
           \hline
        $T$  & $87\%$ & $87\%$ & $80\%$\\
        $P$  & $87\%$ & $86\%$ & $71\%$\\
       $RH$  & $85\%$ & $85\%$ & $80\%$\\
   $T_{DP}$  & $85\%$ & $85\%$ & $80\%$\\
  $w_{dir}$  & $87\%$ & $85\%$ & $75\%$\\
   $w_{sp}$  & $87\%$ & $86\%$ & $80\%$\\
           \hline
        \end{tabular}
        \end{center}
\end{table}

%\begin{table}
%      \caption{Accuracy in the determination of the annual averages of $T$, $P$ and $RH$.}
%    \begin{center}\scriptsize
%   \label{accuracies}
%        \begin{tabular}{ c | c  c  c }
%           \hline
%     Parameter & \multicolumn{3}{c}{Accuracy}\\
%     & Paranal & CAMC & TNG\\
%           \hline
%        $T$  & $0.2^{\circ}$C & $0.6^{\circ}$C & $0.6^{\circ}$C\\
%        $P$  & 0.2 hPa & 0.4 hPa & $-$\\
%       $RH$  & 1\% & 2\% & $-$\\
%           \hline
%        \end{tabular}
%        \end{center}
%\end{table}

\section{Data reduction}\label{data_reduction}
Table \ref{telescopes} that lists positions and
heights of the telescopes at observatories. Paranal is almost 300 m higher than CAMC and
about 4$^{\circ}$ closer to the Equator.\\
Table \ref{databases} reports the databases characteristics.
The Paranal Astronomical Site Monitor is located in the
north area of the Paranal Observatory platform and hosts several
instruments used in the characterization of the site. In particular
in this paper we make use of the meteorological data from the
Vaisala tower.\\
% and seeing data coming from the Paranal Differential Image Motion Monitor (DIMM).\\
%The DIMM is located on a 6 m tower at about 50 m at north-east direction from the Unit Telescope 3 (UT3).
% It delivers seeing measurement with an accuracy better then 10\% above 0.25 arcsec (Sandrock et
% al \cite{sandrock99}).\\
The Vaisala tower is a robust steel structure having a total
height of 30 m. All
the data are regularly collected since January 1st, 1985 and have to
be intended as 20 minutes averages. The external air temperature
($T$) is simultaneously measured at 2 and 30 m above the ground
(same height of the VLT's domes) with an accuracy of $\pm
0.2^{\circ}$C (Sandrock et al.\citealt{sandrock99}).
%giving the possibility to calculate istantaneous vertical temperature gradients to be correlated with seeing data.
Air pressure ($P$) and relative humidity ($RH$) are measured at 2 m above the ground with an accuracy
of $\pm 0.1$ hPa and $\pm 1\%$ respectively.\\
Wind speed ($w_{sp}$, in [m s$^{-1}$]) and wind direction
%($w_{dir}$, N = 0$^{\circ}$, E = 90$^{\circ}$)
 measurements are taken at 10 m above the ground
with a precision of $\pm 0.5$ m s$^{-1}$ for the wind speed and $\pm
3^{\circ}$ for the wind direction. Following the same statistical
procedures of Paper I and Paper II we have computed the hourly
averages and then the monthly averages starting from the $T$, $P$,
$RH$ and $w_{sp}$ raw data series.\\
A particular care is used to minimize any effect due to biases in case of lacking of data 
that typically occurred in winter time. For each missing month value we take into account 
the average obtained from the two corresponding months in other years in which the values 
of the months before and after the absent one are similar. For example, if the lacking month
is September 2002, we look for the two Augusts and Octobers in the other years having 
similar mean values of August and October 2002. The accepted September 2002 value is 
the average of the Septembers corresponding to the chosen Augusts and Octobers. This is 
the main reason why we decided to use monthly averages as an intermediate step in the 
calculation of the annual averages. Finally, we compute the annual averages of T from the 
monthly ones for the three telescopes (see Paper I).\\
We provide to report in tables the annual values and the 
rms of the annual averages of the most important meteo parameters.\\
Wind direction
statistics is evaluated by calculating the annual percentage of
hours in which the wind comes from each direction $\vec{D}$. The wind
rose has been divided into 8 mean directions (N, NE, E, SE, S, SW,
W, NW) and the percentages of hours are calculated into intervals
defined as $[\vec{D}-22.5^{\circ}, \vec{D}+22.5^{\circ}[$.\\
CAMC data from Paper I and Paper II are used in order to compare ORM
with Paranal. For what concerns the temperature trend at ORM we decided to take into account
also data recorded at TNG because they extent to year 2007 (3 years more than CAMC).
Telescopes at ORM are located on a space baseline of about 1000 m (see Paper I).\\
The TNG meteo tower is a robust steel structure with a total height of 15 m. The tower is located about
 100 m far from TNG building. The data are regularly sent from the tower to TNG annex building by 
means of an optic fiber link since 27 March 1998. The data sampling rate is 10 seconds, while data storage
 is done every 30 seconds.\\
The CAMC carried out regular meteorological observations in 
the period 13 May 1984 to 31 March 2005 and the records are more or less continuous in that
period. For the years 1984, 1985 and 1986 meteorological readings are only available at 30 minute intervals. 
From January 1987 readings were made at 5 minute intervals throughout the day and
night regardless of whether observing was in progress. Beginning in December 1994, all readings were made 
at 20 seconds intervals and then averaged over 5 minutes\footnote{http://www.ast.cam.ac.uk}.\\
Both Paranal and ORM are located well above the inversion
layer, in fact the altitude of the inversion layer at Paranal is about 1000 m as reported by Sarazin \cite{sarazin94}, while at ORM it occurs in the range between 800 m and 1200 m (McInnes \& Walker\citealt{mcinnes}).

\begin{table*}%[b]
    \begin{center}
      \caption[]{Mean annual temperatures at Paranal, CAMC and TNG [$^{\circ}$C]. Values in parenthesis correspond to the rms of the annual averages.}
   \label{t-annual-avrg}\scriptsize
        \begin{tabular}{l r r r r r r r r }
           \hline
    Year & 1985 & 1986 & 1987 & 1988 & 1989 & 1990 & 1991 & 1992\\
           \hline
      Paranal $T_{2} $ & 12.2(1.4) & 12.7(1.1) &  $-$ & 12.8(2.0) & 11.9(1.6) & 12.2(1.7) &  $-$ &  $-$\\
      Paranal $T_{30}$ &  $-$ &  $-$ &  $-$ &  $-$ &  $-$ &  $-$ &  $-$ &  $-$\\
     CAMC $T_{10}$ &  8.8(4.4) &  8.9(3.9) &  9.1(3.1) &  7.4(5.4) &  5.2(4.3) &  8.8(5.1) &  8.7(4.8) &  7.9(4.2)\\
      TNG $T_{10}$ &  $-$ &  $-$ &  $-$ &  $-$ &  $-$ &  $-$ &  $-$ &  $-$\\
          \hline
           \hline
    Year & 1993 & 1994 & 1995 & 1996 & 1997 & 1998 & 1999 & 2000\\
           \hline
      Paranal $T_{2} $ & 12.8(1.4) & 13.2(1.3) & 13.1(1.3) & 12.5(1.2) & 13.1(1.6) & 12.9(2.1) & 12.3(1.4) & 12.0(1.7)\\
      Paranal $T_{30}$ &  $-$ &  $-$ &  $-$ &  $-$ &  $-$ &  $-$ &  $-$ & 11.9(1.7)\\
     CAMC $T_{10}$ &  7.0(5.2) &  9.8(4.9) &  9.5(4.3) &  8.6(5.1) &  8.9(4.0) & 10.0(5.0) &  9.3(5.2) &  9.6(5.1)\\
      TNG $T_{10}$ &  $-$ &  $-$ &  $-$ &  $-$ & 10.1(5.1) &  9.6(5.3) &  9.9(5.3)\\
           \hline
           \hline
    Year & 2001 & 2002 & 2003 & 2004 & 2005 & 2006 & 2007 & Average\\
           \hline
      Paranal $T_{2} $ & 12.8(1.3) & 12.8(1.6) & 13.6(1.3) & 12.9(1.5) & 13.0(1.3) & 13.6(0.8) & $-$ & 12.8 $\pm$ 0.5\\
      Paranal $T_{30}$ & 12.6(1.3) & 12.6(1.7) & 13.5(1.3) & 12.7(1.4) & 12.6(1.4) & 13.3(0.9) & $-$ &  $-$\\
     CAMC $T_{10}$ & 10.1(4.3) &  9.6(4.3) &  9.8(4.8) &  9.0(4.9) &  $-$ &  $-$ & $-$ &  8.8 $\pm$ 1.2\\
      TNG $T_{10}$ & 10.7(4.1) &  9.7(4.5) &  9.7(4.8) &  8.9(5.0) &  9.5(4.1) & 10.0(4.7) & 9.6(5.1) &  $-$\\
           \hline
        \end{tabular}
        \end{center}
\end{table*}

\section{Analysis of $T$, $P$, $SOI$ and $RH$}\label{T_P_RH_analysis}

\subsection{Temperature}\label{temperature}
Table \ref{t-annual-avrg} reports the computed mean annual temperatures for Paranal, CAMC and TNG. Values in parenthesis correspond to the rms of the annual averages.\\
The Paranal temperature is taken at two different levels, at 2 and 30 m
above the ground ($T_{2}$ and $T_{30}$ respectively). The vertical
variation of temperature with the altitude (wet and dry adiabatic
lapse rate) is between 6.0 and  10.0$^{\circ}$C/km (Kittel \& Kroemer\citealt{kittel80}), so we expect a
$\sim$$0.2$-$0.3^{\circ}$C
difference in temperature between the two sensors at 2 and 30 m. As
shown in Table \ref{t-annual-avrg}, the mean difference of the
temperature taken at the two heights is 0.2$^{\circ}$C, that is
comparable with the accuracy of the sensor. For this reason we
decide to use the 22 years long database at 2 m as representative of
the temperature at Paranal.
For completeness, Table \ref{t-annual-avrg} reports also the annual $T$ of CAMC measured at 10.5 m above ground
and TNG measured at 10.0 m above ground (see Paper I).\\
\begin{figure}
  \centering
   \includegraphics[width=8cm]{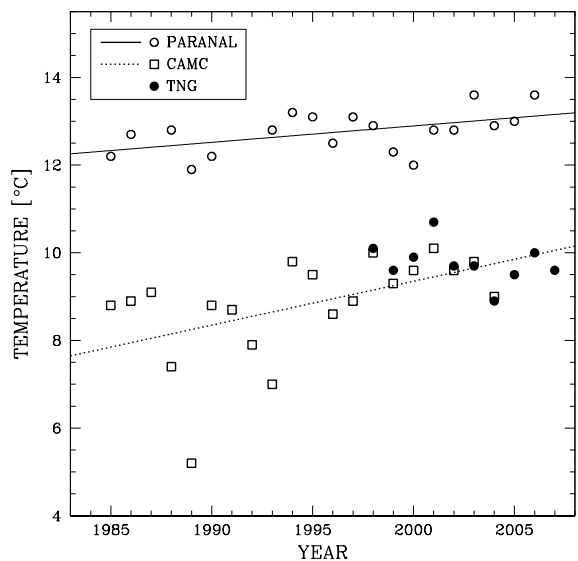}
  \caption{Annual temperatures at Paranal (circles), CAMC (squares) and TNG (filled circles). The solid and the dotted lines
  indicate the linear fit of the data for Paranal and CAMC respectively.}
             \label{t-comparison}
   \end{figure}
\begin{figure}%[b]
  \centering
  \includegraphics[width=8cm]{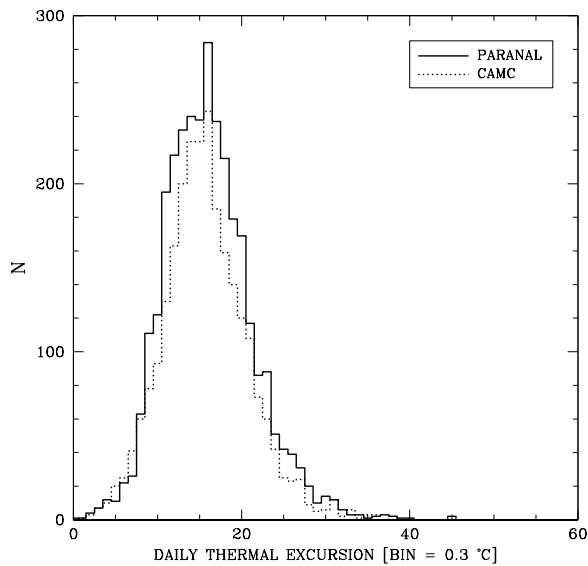}
  \caption{Thermal excursion distribution at Paranal (solid line) and CAMC (dotted line) after 1998. The two sites are almost equivalent.}
             \label{texc}
   \end{figure}
Figure \ref{t-comparison} shows the plot of the $T_{2}$ annual mean values reported in
Table \ref{t-annual-avrg} as well as the annual temperature measured
at CAMC and TNG. The data show an offset of about 4$^{\circ}$C between Paranal and ORM.
Furthemore the rms of the annual averages are about 3 times higher 
at ORM with respect to Paranal. This suggests an higher temperature variations during the 
years at ORM.
Trends at Paranal and CAMC show a positive slope during the years, while
TNG data have a flat trend. The
best linear fit of CAMC data gives an increase of the temperatures
of about $(1.0 \pm 0.3)$$^{\circ}$C/10yrs, while the slope computed for Paranal
gives a value of $(0.4 \pm 0.1)$$^{\circ}$C/10yrs. This is not surprising considering
that the global warming and the glacier retreat are less pronounced in South America.
To facilitate the comparison, it is also drawn the total average for
Paranal and CAMC.\\
But it is interesting to note that after 1998 data at TNG seem to indicate a flattening of
the increasing temperature trend. This rises the question if we are in presence of a global
warming or it is a typical temperature oscillation through decades and/or a regional effect.\\
The 22 years long baseline of Paranal is characterized
by an increasing trend, the linear regression of Paranal data in the
period 1993-2000 shows a slope of $(-0.12\pm0.05)$ while data in the period
2000-2006 have an opposite slope slightly steeper $(0.19\pm0.07)$. This
suggests a possible correlation with the occurrence of wide-scale
climatological events such as El Ni\~no and La Ni\~na phenomenons. In
fact, a preliminary check has shown the presence of strong
La Ni\~na episodes in 1999 and 2000 that are linked to the coldest annual temperatures at Paranal
(see Section \ref{SOI}).\\
This confirms the strong link between air
and ocean temperature that may influences the high level of the
atmosphere.\\
The year 1989 appears to be the coldest for both sites in the last 20 years. The year
2001 is the warmest for CAMC while the 2003 and 2006 are the warmest for Paranal.\\
The typical 3$-$4 years oscillation of the temperatures at CAMC pointed in Paper I is not so clear for Paranal.\\
An interesting results comes from the analysis of the daily thermal excursion after 1998
in the two sites. in fact looking at Figure \ref{texc} we do see a very similar distribution for both sites with a peak
at bin number 18, corresponing to a thermal excursion of about 4.8$^{\circ}$C.
\begin{figure}%[t]
  \centering
 \includegraphics[width=8cm]{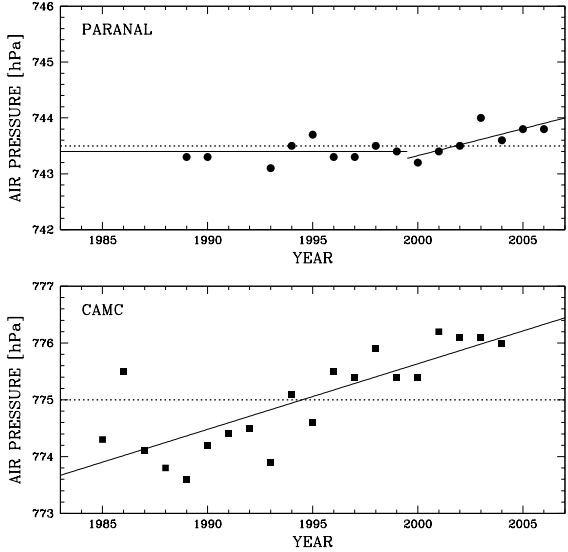}
  \caption{$Top$: annual $P$ at Paranal Observatory. $Bottom$: annual $P$ at CAMC. The averages are represented by the dotted lines, while the linear fits by the solid lines.}
             \label{p_comparison}
   \end{figure}
\begin{table*}%[b]
    \begin{center}
      \caption[]{Mean annual air pressure at Paranal and CAMC [hPa]. Values in parenthesis correspond to the rms of the annual averages.}
   \label{p-annual-avrg}\scriptsize
        \begin{tabular}{ l r r r r r r r r }
           \hline
    Year & 1985 & 1986 & 1987 & 1988 & 1989 & 1990\\
           \hline
      Paranal &   $-$ &   $-$ &   $-$ &   $-$ & 743.3(1.2) & 743.3(0.9)\\
     CAMC & 774.3(2.6) & 775.5(2.3) & 774.1(1.8) & 773.8(2.2) & 773.6(2.3) & 774.2(2.4)\\
           \hline
           \hline
    Year  & 1991 & 1992 & 1993 & 1994 & 1995 & 1996\\
           \hline
      Paranal &   $-$ &   $-$ & 743.1(0.7) & 743.5(0.5) & 743.7(0.7) & 743.3(0.6)\\
     CAMC & 774.4(2.5) & 774.5(2.2) & 773.9(2.7) & 775.1(1.8) & 774.6(2.2) & 775.5(3.8)\\
           \hline
           \hline
    Year & 1997 & 1998 & 1999 & 2000 & 2001 & 2002\\
           \hline
      Paranal & 743.3(0.7) & 743.5(0.5) & 743.5(0.4) & 743.2(0.6) & 743.4(0.4) & 743.5(0.3)\\
     CAMC & 775.4(2.4) & 775.9(2.2) & 775.4(2.2) & 775.4(2.4) & 776.2(2.5) & 776.1(2.4)\\
           \hline
           \hline
    Year & 2003 & 2004 & 2005 & 2006 & & Average\\
           \hline
      Paranal & 743.0(0.4) & 743.6(0.4) & 743.8(0.4) & 743.8(0.4) & & 743.5 $\pm$ 0.2\\
     CAMC & 776.1(2.4) & 776.0(2.7) &   $-$ &   $-$ & & 775.0 $\pm$ 1.0\\
           \hline
        \end{tabular}
        \end{center}
\end{table*}
\subsection{Air pressure}\label{pressure}
\begin{figure}%[t]
  \centering
 \includegraphics[width=8cm]{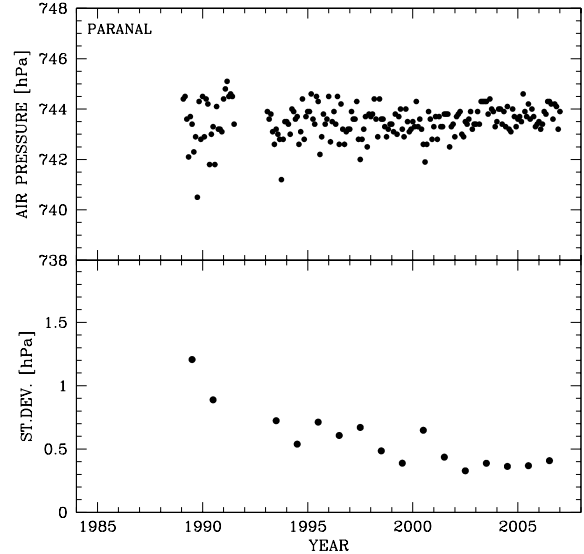}
  \caption{$Top$: Monthly $P$ at Paranal. $Bottom$: Standard deviation of the monthly $P$ at Paranal.}
             \label{p-pressure_monthly}
   \end{figure}

\begin{figure}%[t]
  \centering
  \includegraphics[width=8cm]{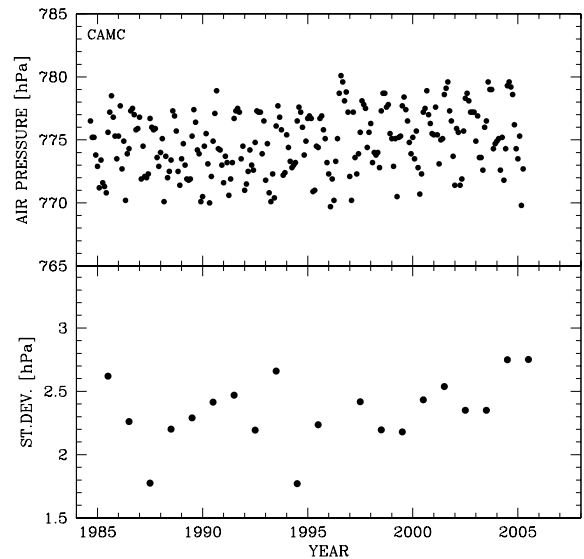}
  \caption{$Top$: Monthly $P$ at CAMC. $Bottom$: Standard deviation of the monthly $P$ at CAMC.}
             \label{o-pressure_monthly}
   \end{figure}

Figure \ref{p_comparison} shows the annual mean pressure  $P$
calculated at Paranal and CAMC. Table \ref{p-annual-avrg} reports
the values plotted in the figure. The average pressure at Paranal
during the entire baseline is 743.5 hPa, while the average pressure
at CAMC is 775.0 hPa. The plot shows two different behaviors at
Paranal with a changing of slope in the year 2000. Splitting the fit
in two subsamples we see that the best fit of the points before the
2000 is almost flat ($\pm$0.1 hPa) with a mean value of 743.4 hPa. The best fit
after the year 2000 is characterized by a positive slope of about
$(0.1 \pm 0.1)$ hPa and the mean pressure is 743.5 hPa. This difference is
comparable to the accuracy of the measuring instruments. A more
detailed analysis is needed to check if this is a long term
increasing trend of pressure.
The data obtained from CAMC show a similar increasing trend during the years ($\sim$$1.5$ hPa, see Paper II).\\
The minimum pressure  measured at Paranal was 728.0 hPa in September
1989, while the highest one was in April 1998 with a value of 754.0 hPa.
Instead, CAMC minimum pressure was 747.7 hPa in September 2004, while the maximum was 785.8 hPa in July 2001.\\
Two interesting plots are shown in Figures \ref{p-pressure_monthly} and \ref{o-pressure_monthly},
where the monthly $P$ averages in the two sites are plotted.
There is a clear decreasing of the dispersion of the monthly $P$ during the
years at Paranal (Figure \ref{p-pressure_monthly}, $top$) as confirmed by the analysis of the standard deviation
of the annual averages (Figure \ref{p-pressure_monthly}, $bottom$).
In fact at Paranal it decreased of about 70\% between 1990 and 2006,
while the same data computed at CAMC do not show the same phenomenon (Figure \ref{o-pressure_monthly}).
The same analysis applied to monthly temperatures shows a lower seasonal dispersion at Paranal ($\sim$$8^{\circ}$C)
with respect to CAMC ($\sim$$19^{\circ}$C).
Temperature dispersion remained almost constant up to now in both sites.\\
To check if the effect of the yearly variation of the pressure at
Paranal can be explained taking into account wide-scale
climatological episodes, we have checked  a possible correlation
with the Souther Oscillation Index. This analysis will
be discussed in Section \ref{SOI}.\\
Following Paper II, we have calculated the theoretical pressure for
Paranal using the barometric correction that depends on site's scale
height $H$ in the barometric law. The knowledge of the average pressure and the atmospheric scale
height are important parameters to compute the local contribution of
the Rayleigh scattering to the total astronomical extinction (Hayes \& Latham\citealt{hayes75}).\\
At CAMC we have found a typical
scale height $H_{CAMC} = 8325$ m and a theoretical pressure of 766.0
hPa in the period between 1998-2004, while the data give 775.9
hPa on average, confirming that ORM is dominated by high pressure (Paper II,
Section 5.1). Since the barometric correction is a function of the
mean temperature $\left\langle T_{layer}\right\rangle$ of the layer
in which the correction is applied, we have taken into account as
closest weather station the city of Antofagasta, close to the
Pacific Coast, few meters above sea level. Between 1998 and 2004 the
mean temperature of Antofagasta was $16.7 \pm 0.4^{\circ}$C
\footnote{See http://www.tutiempo.net.}, while
the mean temperature computed at Paranal in the same years was $12.8
\pm 0.5^{\circ}$C.
These two  mean temperatures (Paranal and Antofagasta) allow us to
calculate the mean temperature of the layer between the sea level and the Paranal,
using the weighted average of the two measurements.
We obtain $\left\langle T_{layer}\right\rangle = 15.2 \pm 0.5^{\circ}$C.\\
The standard atmospheric model
from NASA's Glenn Research Center (GRC) Web site\footnote{See
http://www.grc.nasa.gov.} permits us to calculate the theoretical
pressure at the altitude of Paranal for a $\left\langle
T_{layer}\right\rangle = 15.0^{\circ}$C which gives a resulting
theoretical pressure of 734.4 hPa, and a theoretical scale height
$H_{GRC} \approx 8238$ m. Instead, using standard tables in Allen
\cite{allen00}, for $\left\langle T_{layer}\right\rangle =
15.0^{\circ}$C, we find a theoretical scale height $H_{Allen}
\approx 8430$ m, in good agreement with $H_{GRC}$. The mean value
between $H_{GRC}$ and $H_{Allen}$ gives us $H_{Paranal} = 8334$ m which
corresponds to a theoretical pressure for Paranal of $P_{Paranal} =
738.3$ hPa. This result is lower with respect to the empirical
pressures reported in Table \ref{p-annual-avrg}, so we can confirm
that also Paranal is dominated by high pressure. The Paranal scale
height is a bit higher than at ORM (8334 m vs 8325 m) as expected
having a lower latitude site and higher average temperature than
ORM.
  \begin{table*}%[t]
    \begin{center}
      \caption[]{Seasons definition at Paranal and CAMC.}
   \label{dew_seasons}\scriptsize
        \begin{tabular}{ r | l | l }
           \hline
    & Paranal & CAMC\\
           \hline
    WINTER & July-August-September & January-February-March\\
    SPRING & October-November-December & April-May-June\\
    SUMMER & January-February-March & July-August-September\\
    AUTUMN & April-May-June & October-November-December\\
           \hline
        \end{tabular}
        \end{center}
\end{table*}
 \begin{figure}%[t]
  \centering
 \includegraphics[width=8cm]{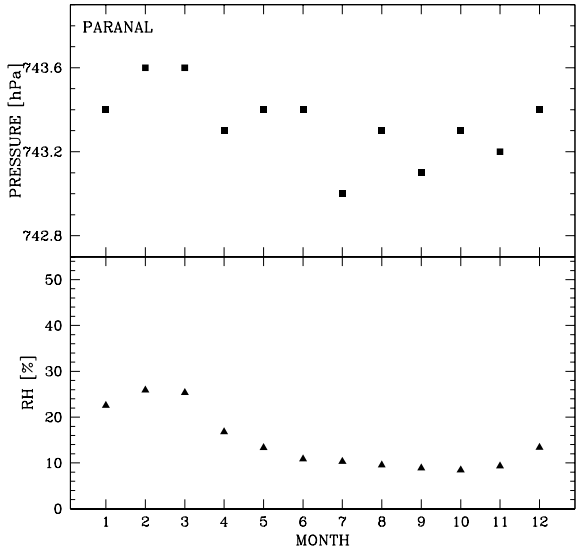}
  \caption{Averages of monthly $P$ ($top$) and $RH$ ($bottom$) at Paranal.}
             \label{prh_paranal}
\end{figure}
\begin{figure}%[t]
  \centering
  \includegraphics[width=8cm]{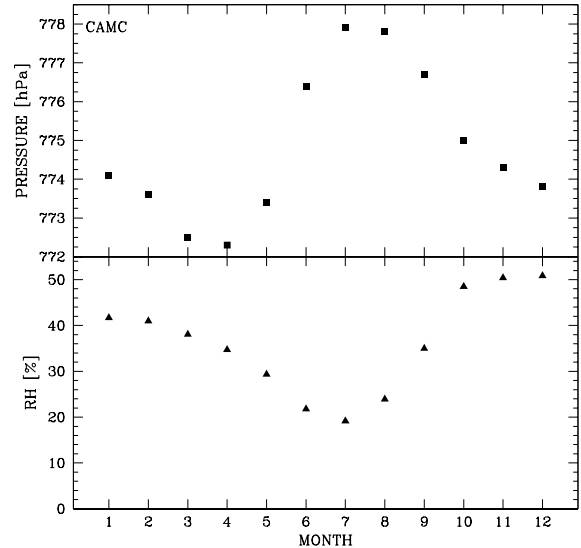}
  \caption{Averages of monthly $P$ ($top$) and $RH$ ($bottom$) at CAMC.}
             \label{prh_camc}
\end{figure}

\subsection{Relative humidity}\label{rh}
The relative humidity ($RH$) and the dew point are two important
parameters for the astronomical instrumentation, because they set
the occurrence of moist and water condensation on the coldest part
of the telescope and of the instruments. In particular these
parameters may affect the upper surface of the main mirror and the
pipes of the cooling system.\\
We separated the annual $RH$ in 4 seasons as defined in
Table \ref{dew_seasons}. Note that Paranal and ORM are in
opposite hemispheres, so the definitions are inverted for the two observatories.\\
To facilitate the discussion we consider as \textit{cold season} autumn and winter,
and \textit{warm season} spring and summer.
Tables \ref{rh-winter} and \ref{rh-summer} report the computed annual
$RH$ in cold and warm seasons at Paranal and CAMC.
Typically in cold seasons $RH$ is lower than 60\% at CAMC and $15\%$ at Paranal.
In warm seasons $RH$ is lower than 40\% at CAMC and $20\%$ at Paranal.
Figures \ref{prh_paranal} and \ref{prh_camc} show the seasonal trend of $P$ ($top$) and $RH$ ($bottom$) in both sites
considered as the average of the same month through the considered years of the databases.
The variation of $P$ at Paranal is $< 1$ hPa, very lower with respect to CAMC ($\sim$$6$ hPa).
Variations of pressure in a short time scale (few hours) can induce weather instabilities,
while in a long time scale they increase the differences in the weather in different seasons.\\
We clearly see an anti-correlation between $P$ and $RH$ in summer at CAMC, while at Paranal this effect is not obvious.
Because of the effect of the Bolivian Winter, driving equatorial humid air from
amazonian basin along the Andes, the $RH$ at Paranal is higher in warm seasons
($\sim$$20\%$) with respect to cold ones ($\sim$$12\%$).
Higher $RH$ appears clearly in wintertime than in summertime at CAMC.\\
\begin{table*}%[b]
    \begin{center}
      \caption[]{Mean annual $RH$ in cold seasons at Paranal and CAMC [\%]. Values in parenthesis correspond to the rms of the annual averages.}
   \label{rh-winter}\scriptsize
        \begin{tabular}{l r r r r r r r r }
           \hline
    Year & 1985 & 1986 & 1987 & 1988 & 1989 & 1990 & 1991 & 1992\\
           \hline
     Paranal &  9.1(2.4) &  9.7(4.6) &  $-$ &  9.2(5.1) & 12.0(5.5) &  9.6(2.1) &  $-$ &  $-$\\
    CAMC & 40.3(7.0) & 43.8(4.1) & 48.2(2.5) & 52.2(7.0) & 49.2(3.6) & 56.5(14.7) & 57.4(10.2) & 52.7(4.7)\\
           \hline
           \hline
    Year & 1993 & 1994 & 1995 & 1996 & 1997 & 1998 & 1999 & 2000\\
           \hline
     Paranal & 10.8(3.8) & 11.0(4.2) & 10.5(3.1) & 10.7(3.0) & 14.8(3.9) & 12.6(4.3) & 10.9(2.0) & 14.9(4.7)\\
    CAMC & 47.8(9.3) & 40.4(8.8) & 40.6(8.6) & 53.1(12.7) & 44.7(12.4) & 30.9(6.1) & 12.5(7.7) & 53.0(13.6)\\
           \hline
           \hline
    Year & 2001 & 2002 & 2003 & 2004 & 2005 & 2006 & & Average\\
           \hline
     Paranal & 12.5(4.3) & 14.5(6.3) & 11.4(3.0) & 12.4(3.0) & 11.5(1.3) & 12.4(3.6) & & 11.6 $\pm$ 1.8\\
    CAMC & 33.7(14.7) & 56.4(8.9) & 36.3(10.8) & 40.8(12.8) &  $-$ &  $-$ & & 44.5 $\pm$ 10.1\\
           \hline
        \end{tabular}
        \end{center}
\end{table*}
\begin{table*}%[b]
    \begin{center}
      \caption[]{Mean annual $RH$ in warm seasons at Paranal and CAMC [\%]. Values in parenthesis correspond to the rms of the annual averages.}
   \label{rh-summer}\scriptsize
        \begin{tabular}{l r r r r r r r r }
           \hline
    Year & 1985 & 1986 & 1987 & 1988 & 1989 & 1990 & 1991 & 1992\\
           \hline
     Paranal &  $-$ & 16.9(8.3) & 15.8(7.0) & 13.0(4.6) & 16.9(12.9) & 11.5(4.4) & 16.3(6.8) &  $-$\\
    CAMC & 43.1(3.5) & 42.0(7.3) & 43.6(4.6) & 30.8(7.8) & 35.9(10.1) & 42.5(14.3) & 29.2(10.3) & 26.1(12.0)\\
           \hline
           \hline
    Year & 1993 & 1994 & 1995 & 1996 & 1997 & 1998 & 1999 & 2000\\
           \hline
     Paranal &  $-$ & 15.8(5.7) & 16.2(6.3) & 15.0(7.3) & 20.0(12.2) & 16.4(7.6) & 21.1(12.2) & 25.5(14.4)\\
    CAMC & 21.0(14.1) & 16.7(11.1) & 26.5(10.8) & 15.0(6.2) & 15.6(15.0) &  6.5(5.1) & 19.7(10.2) & 24.3(13.6)\\
           \hline
           \hline
    Year & 2001 & 2002 & 2003 & 2004 & 2005 & 2006 & & Average\\
           \hline
     Paranal & 22.8(15.4) & 19.8(11.1) & 15.3(8.2) & 17.0(8.6) & 18.0(6.0) &  $-$ & & 17.4 $\pm$ 3.4\\
    CAMC & 21.9(9.1) & 22.0(15.5) & 24.5(3.6) & 32.2(9.8) &  $-$ &  $-$ & & 27.0 $\pm$ 10.3\\
           \hline
        \end{tabular}
        \end{center}
\end{table*}
A standard requirement for the use of telescopes is a $RH$ value $< 80\%$ or $< 85\%$.
For both sites we have calculated the number of nights in which $RH$ has been higher than the mentioned limits
for more than 50\% of the duration of the night. Only nights which duration has been $\geq 6$ hours
have been used in the calculation. Results are reported in Table \ref{rh_80-85} and denote a
significant difference between Paranal and CAMC, the first appearing almost immune to high $RH$ events.
The number of nights at CAMC is the same for the two imposed limits.
\begin{table}%[t]
    \begin{center}
      \caption[]{Paranal and CAMC: annual number of nights in which $RH$ has been higher than $80\%$ and $85\%$
      for more than 50\% of the duration of the night. Only nights which duration has been $\geq 6$ hours have been
      used in the calculation. For each year, the total number of nights having duration $\geq 6$ is also reported.}
   \label{rh_80-85}\scriptsize
        \begin{tabular}{c | c c | c c | c c}
           \hline
     & \multicolumn{2}{c}{$RH > 80\%$} & \multicolumn{2}{c}{$RH > 85\%$} & \multicolumn{2}{c}{Nights}\\
    Year & Paranal & CAMC & Paranal & CAMC & Paranal & CAMC\\
           \hline
    1984 & $-$ & 2 & $-$ & 2 & $-$ & 155\\
    1985 & 1 & 8 & 0 & 8   & 324 & 253\\
    1986 & 0 & 8 & 0 & 8   & 311 & 285\\
    1987 & 0 & 22 & 0 & 22 & 108 & 332\\
    1988 & 0 & 51 & 0 & 51 & 258 & 349\\
    1989 & 1 & 68 & 0 & 68 & 310 & 362\\
    1990 & 0 & 73 & 0 & 73 & 329 & 354\\
    1991 & 0 & 53 & 0 & 53 & 177 & 358\\
    1992 & $-$ & 58 & $-$ & 58 & $-$ & 366\\
    1993 & 0 & 58 & 0 & 58 & 300 & 363\\
    1994 & 0 & 41 & 0 & 41 & 328 & 355\\
    1995 & 0 & 51 & 0 & 51 & 333 & 355\\
    1996 & 0 & 81 & 0 & 81 & 348 & 354\\
    1997 & 2 & 52 & 0 & 52 & 336 & 364\\
    1998 & 0 & 36 & 0 & 36 & 348 & 359\\
    1999 & 2 & 48 & 2 & 48 & 360 & 354\\
    2000 & 4 & 49 & 2 & 49 & 365 & 333\\
    2001 & 0 & 60 & 0 & 60 & 364 & 353\\
    2002 & 4 & 31 & 3 & 31 & 362 & 274\\
    2003 & 0 & 27 & 0 & 27 & 352 & 346\\
    2004 & 0 & 35 & 0 & 35 & 365 & 334\\
    2005 & 3 & $-$ & 3 & $-$ & 233 & $-$\\
           \hline
        \end{tabular}
        \end{center}
\end{table}
\begin{table}%[b]
    \begin{center}
      \caption[]{Strongest El Ni\~no and La Ni\~na episodes between 1985 and 2006.}
   \label{soi_episodes}\scriptsize
        \begin{tabular}{ r | r }
          \hline
 El Ni\~no & La Ni\~na\\
           \hline
      1987 & 1988\\
 1992-1994 & 1996\\
      1997 & 1999-2000\\
           \hline
        \end{tabular}
        \end{center}
\end{table}
  \begin{figure}%[t]
  \centering
 \includegraphics[width=8cm]{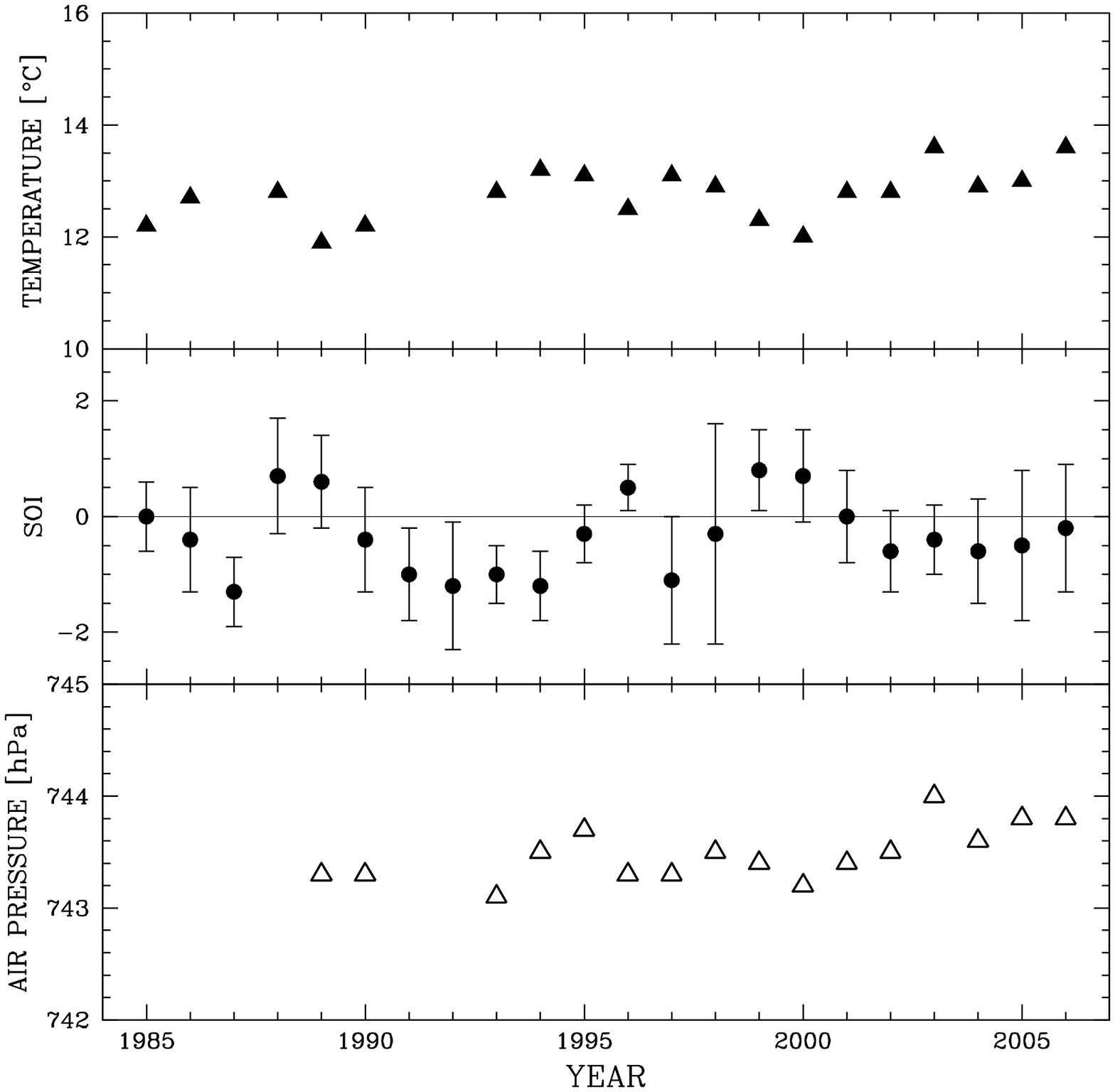}
  \caption{Southern Oscillation Index ($middle$) vs. annual temperatures ($top$) and annual air pressure ($bottom$) at Paranal. The error-bars in the $middle$ plot correspond to the rms of the SOI annual means.}
             \label{soi_comparison}
   \end{figure}
\subsection{The Southern Oscillation Index (SOI)}\label{SOI}
In Paper I we demonstrated the correlation between the North
Atlantic Oscillation Index and the annual mean temperatures at CAMC
(c.l. $\sim$$0.9$). The NAO is the dominant mode of atmospheric
circulation in the North Atlantic region and is generally defined as
the difference in pressure between the Azores high pressure and the
Icelandic low pressure (Wanner et al.\citealt{wanner01}; Graham\citealt{graham05}; Paper I Section 4).
We pointed that the action of a positive NAO
Index is as a brake for the increase in temperatures, and an
accelerator for the decrease. Conversely, a negative NAO Index acts in the opposite manner (see Paper I, Section 4).\\
In the case of Paranal we have analysed the Southern Oscillation
Index (SOI). The SOI is defined as the difference in air pressure
occurring between the western and eastern tropical Pacific (Tahiti
and Darwin, Australia). Changes in the SOI correspond also with
changes in temperatures across the eastern tropical Pacific. The
negative phase of the SOI represents below-normal air pressure at
Tahiti and above-normal air pressure at Darwin (Halpert \& Ropelewski\citealt{halpert92}). Prolonged periods of
negative SOI values reflect abnormally warm ocean waters across the
eastern tropical Pacific, typical of El Ni\~no episodes, while
periods of positive SOI values coincide with abnormally cold ocean
waters across the eastern tropical Pacific,
typical of La Ni\~na episodes\footnote{See http://www.cpc.ncep.noaa.gov.} (Higgins et al.\citealt{higgins01}).\\
Because of the strong influence of the SOI on the meteorological
conditions in the southern hemisphere, and on observing conditions
at Paranal and La Silla,  it is important to investigate possible
correlations between the SOI and the temperature and pressure at
Paranal.
For this reason we calculated the annual averages of the SOI from the monthly averages retrieved from the National Prediction Center Web site.\\
Figure \ref{soi_comparison} ($middle$) shows the SOI as a function
of the year where the error bars correspond to the rms of the SOI annual means.
Each minimum in the figure corresponds to warm episode
(El Ni\~no), while each maximum to cold episode (La Ni\~na). As
shown in the figure and also reported in Table \ref{soi_episodes},
strong El Ni\~no occurred in the years 1987, 1992-1994 and 1997, while strong La Ni\~na occurred in 1988, 1996 and 1999-2000.\\
The $top$ of Figure \ref{soi_comparison} shows the $T$ annual
averages at Paranal. We see an increase in the annual temperatures
in connection with El Ni\~no, vice versa minimum temperatures occur
with La Ni\~na episodes (correlation is $\sim$$0.7$). Local minimums in the $P$ annual trends
at Paranal (Figure \ref{soi_comparison}, $bottom$) are linked to the
presence of La Ni\~na events, while maximums with El Ni\~no, but in
this case poor correlations is found between the yearly trends at
Paranal.
\section{Analysis of the dew point}\label{dewpoint}
The dew point temperature ($T_{DP}$) is the critical temperature
at which condensation occurs.
When dew point temperature and air temperature $T$ are equal
the air said to be saturated and condensation appears if the air cools. 
It is clear that the knowledge of the 
dew point temperature is crucial for the maintenance of the
optics of the telescope to avoid condensation,
in particular if the the instrumentation is maintained at temperatures
few degrees lower than the air temperature.
Condensation can be reached mostly in clear nights, 
when the earth cools rapidly,
therefore it is important to know the percentage of time in which condensation may occur to 
have the best performances from ground based telescopes.
To this aim we have computed the dew point temperature using nighttime data
for both Paranal and CAMC. We have defined as nighttime the range
20:00$-$6:00 hr Local Time. Only clear nights have been taken into account.
The declared percentage of clear nights at ORM is 84\% (Mu\~noz-Tu\~n\'on et al.\citealt{munoz-tunon07}),
while at Paranal it is 85\% (http://www.eso.org).\\
The analysis concerns the annual percentage of time in which $\Delta T = T - T_{DP} < X$,
where $X$ corresponds to a variable upper limit for $\Delta T$ ($X = 1$ and $5^{\circ}$C).
This statistics is crucial for the knowledge of the amount of time in
which a danger of condensation on the telescopes hardware may occur in two extreme $\Delta T$.\\
In Figure 9 we report the computed frequencies in the two sites.
In winter, spring and autumn we see a significant difference
between Paranal and CAMC until 2000. After such year the two sites seem to became more similar.\\
Paranal shows negligible percentages for
all $X$ limits in spring, while it never goes above $\sim$$4\%$ for
$X = 5^{\circ}$C in winter (2002) and autumn (2002 and 2004).
CAMC has frequency often $> 10\%$ when $X = 5^{\circ}$C.
After 2001 CAMC shows similar percentages with
respect to Paranal when $X = 1^{\circ}$C. This is due to the very
little increasing of the percentages at Paranal after 2001.\\
In summer the two sites are very similar maintaining percentages
never higher than $\sim$$3\%$ at Paranal and $\sim$$4\%$ at CAMC
when $X = 5^{\circ}$C, with the exception of the year 1990 for CAMC.
\begin{figure*}%[t]
    \begin{center}
   \label{dewpoint_figures}
\resizebox{0.95\hsize}{!}{
	\includegraphics[width=8cm]{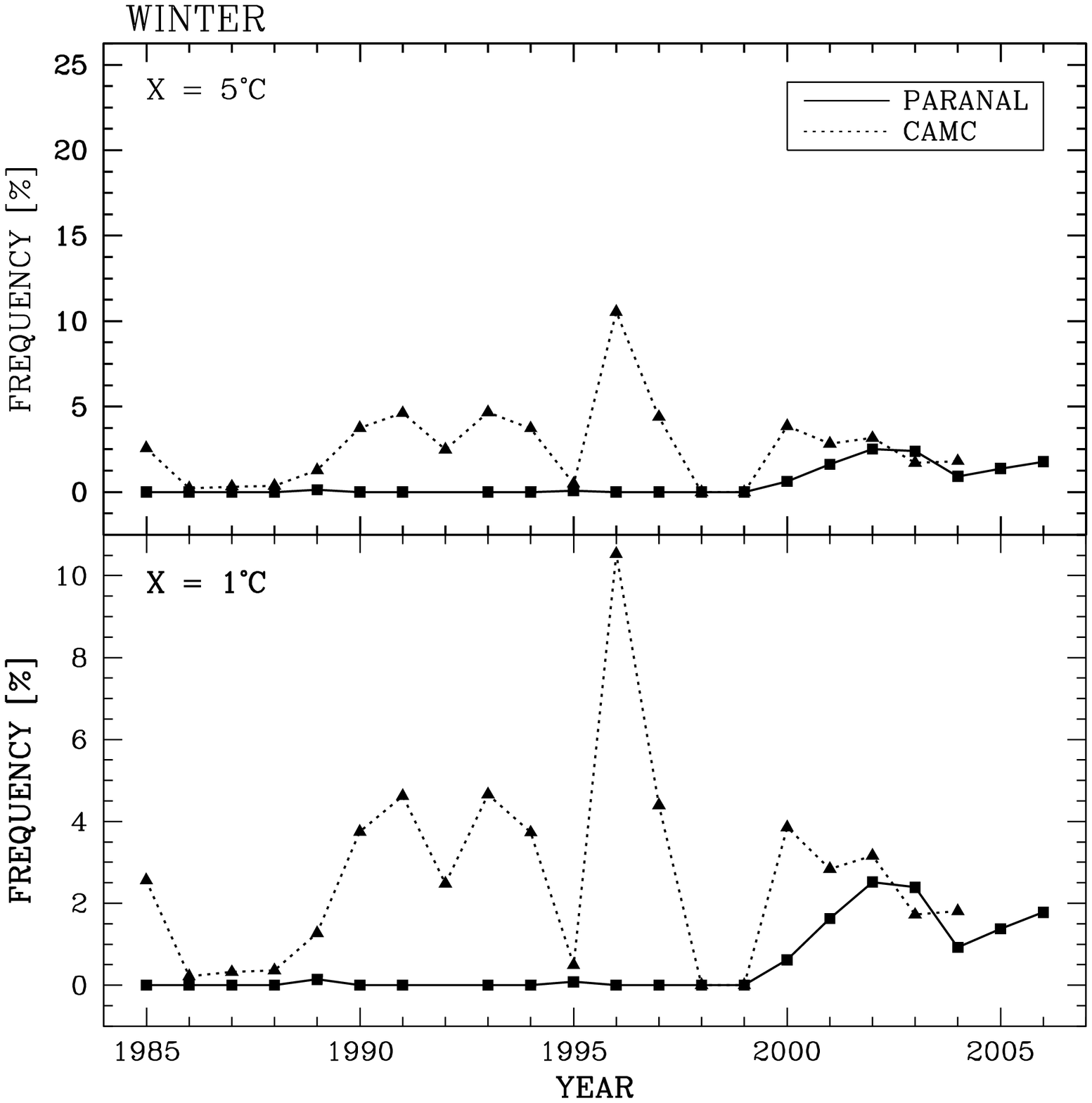}
  \includegraphics[width=8cm]{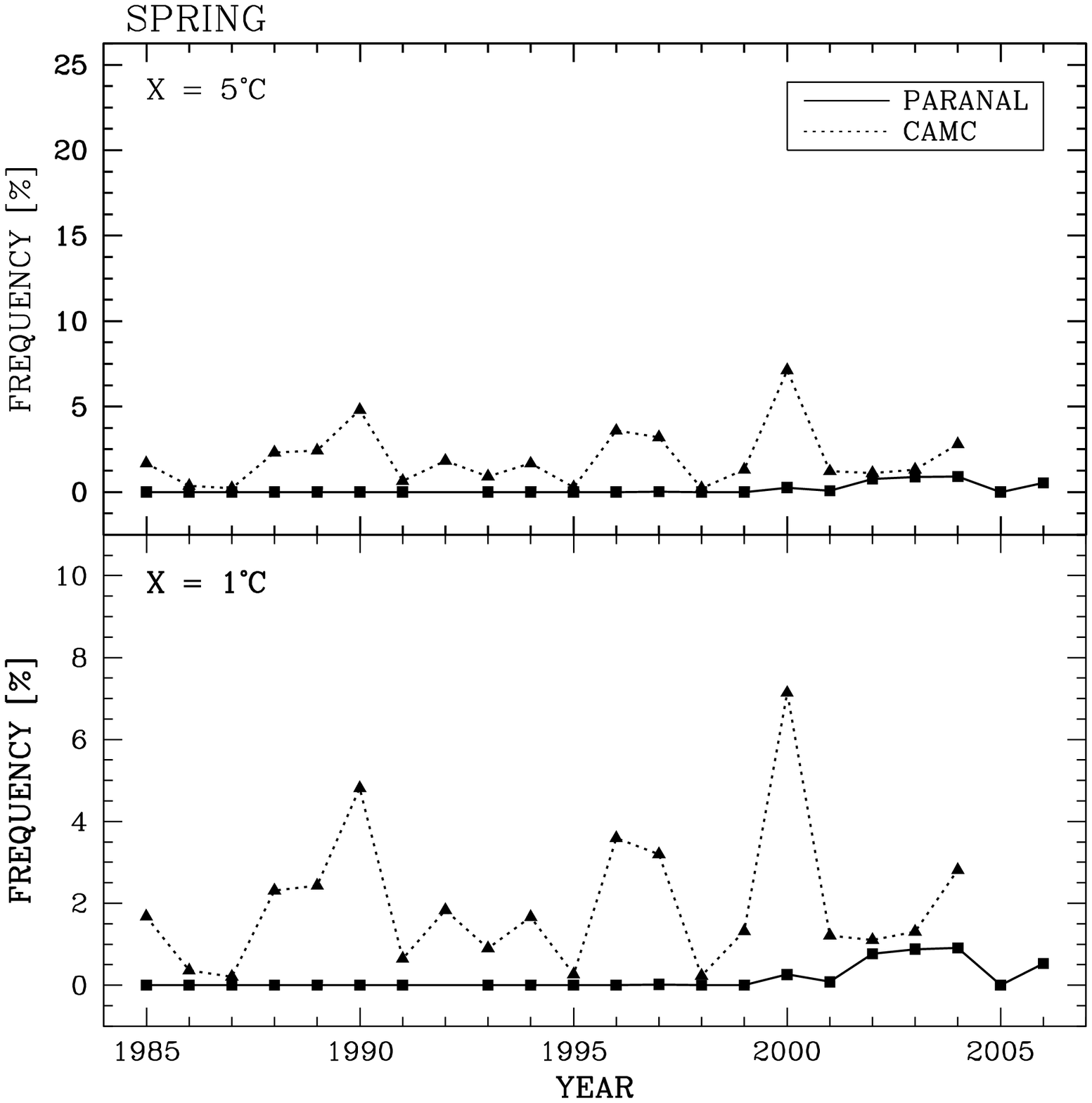}
											}
\resizebox{0.95\hsize}{!}{
  \includegraphics[width=8cm]{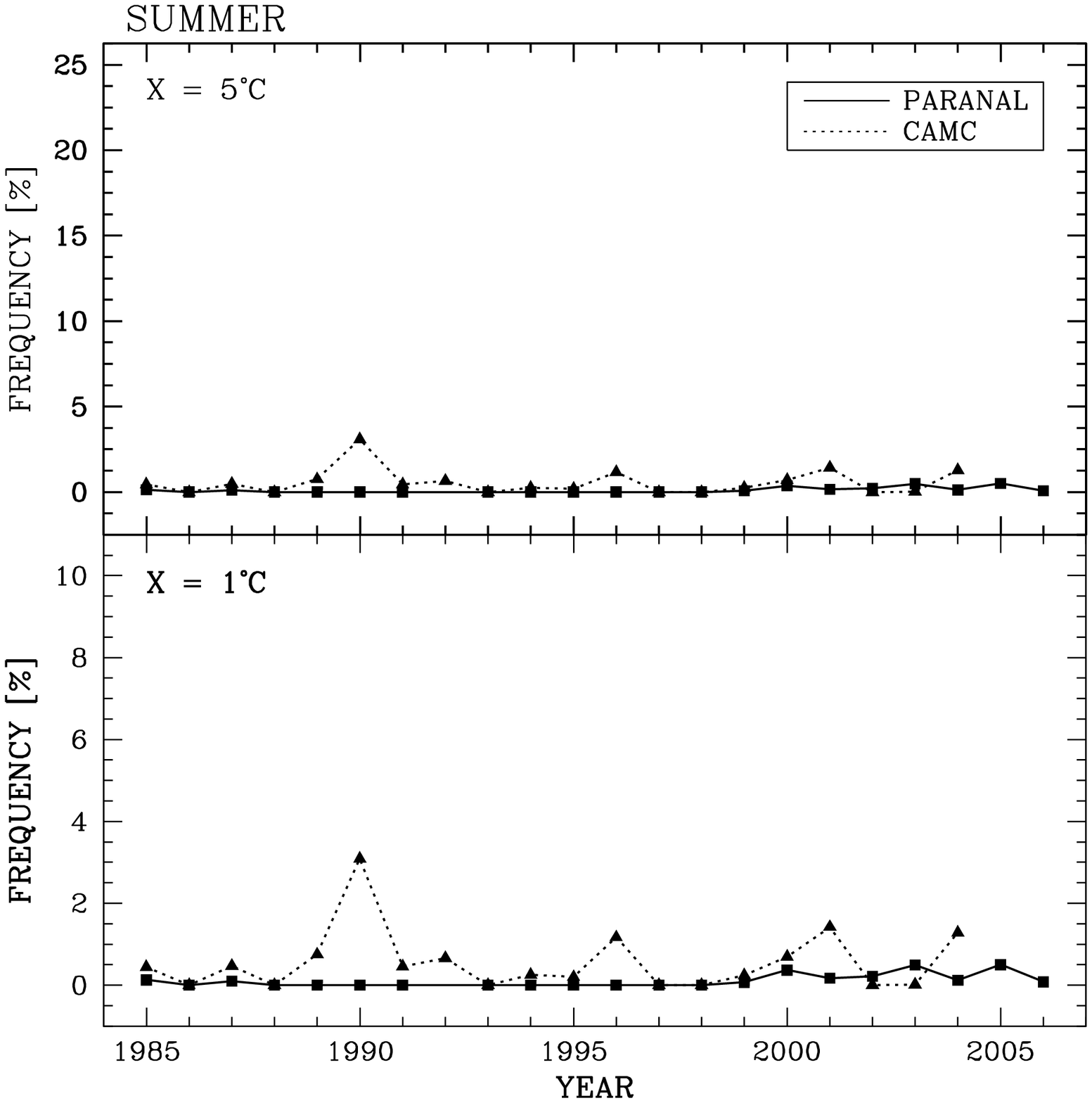}
  \includegraphics[width=8cm]{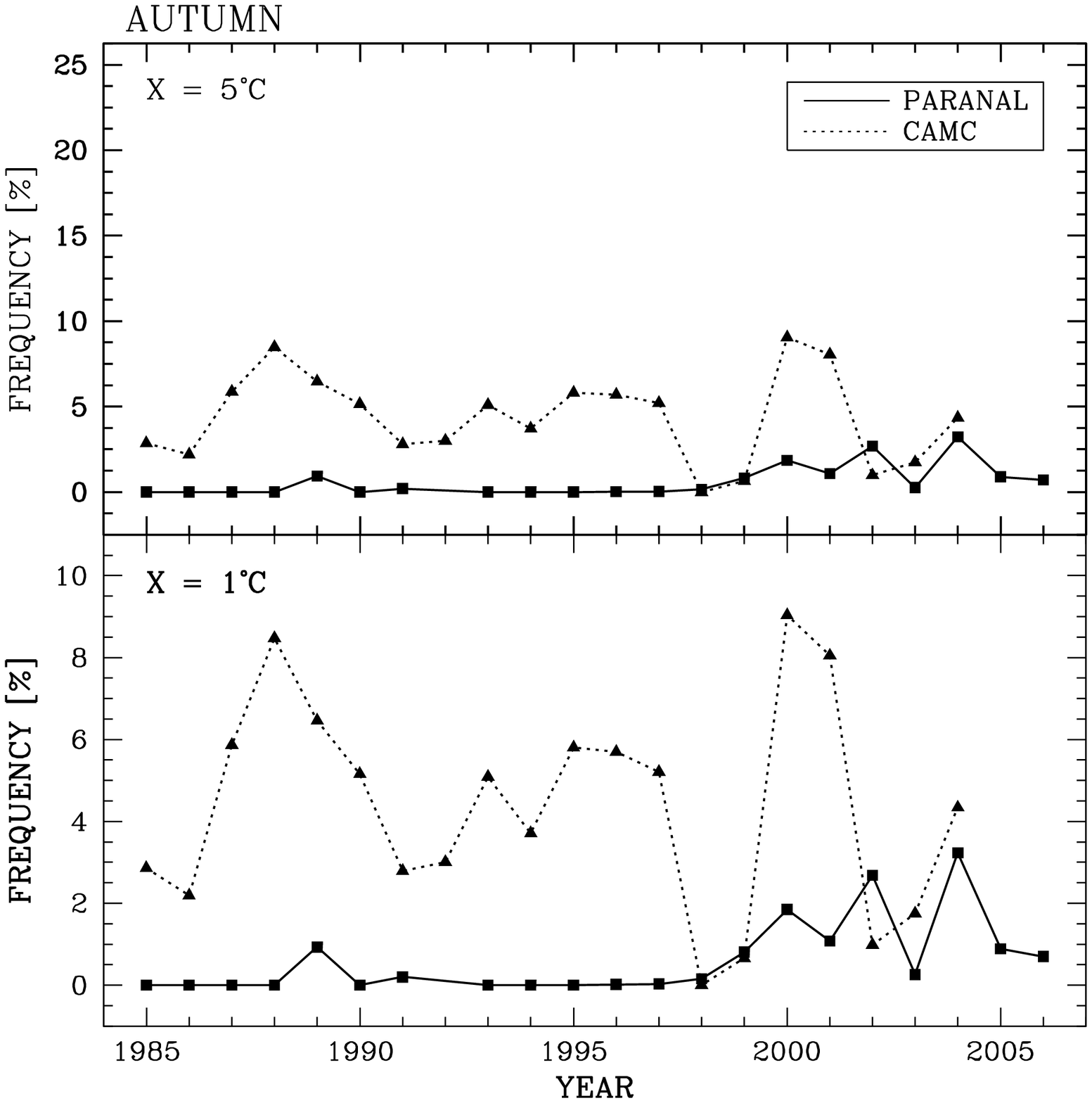}
											}
      \caption{Nighttime annual percentage of time in which $\Delta T = T - T_{DP} < 1$ and $5^{\circ}$C at Paranal and CAMC in winter, spring, summer and autumn.}
        \end{center}
\end{figure*}
\section{Analysis of the wind}\label{wind}
\subsection{Wind direction}\label{w_dir}
\begin{figure}%[t]
\centering
\includegraphics[width=4cm]{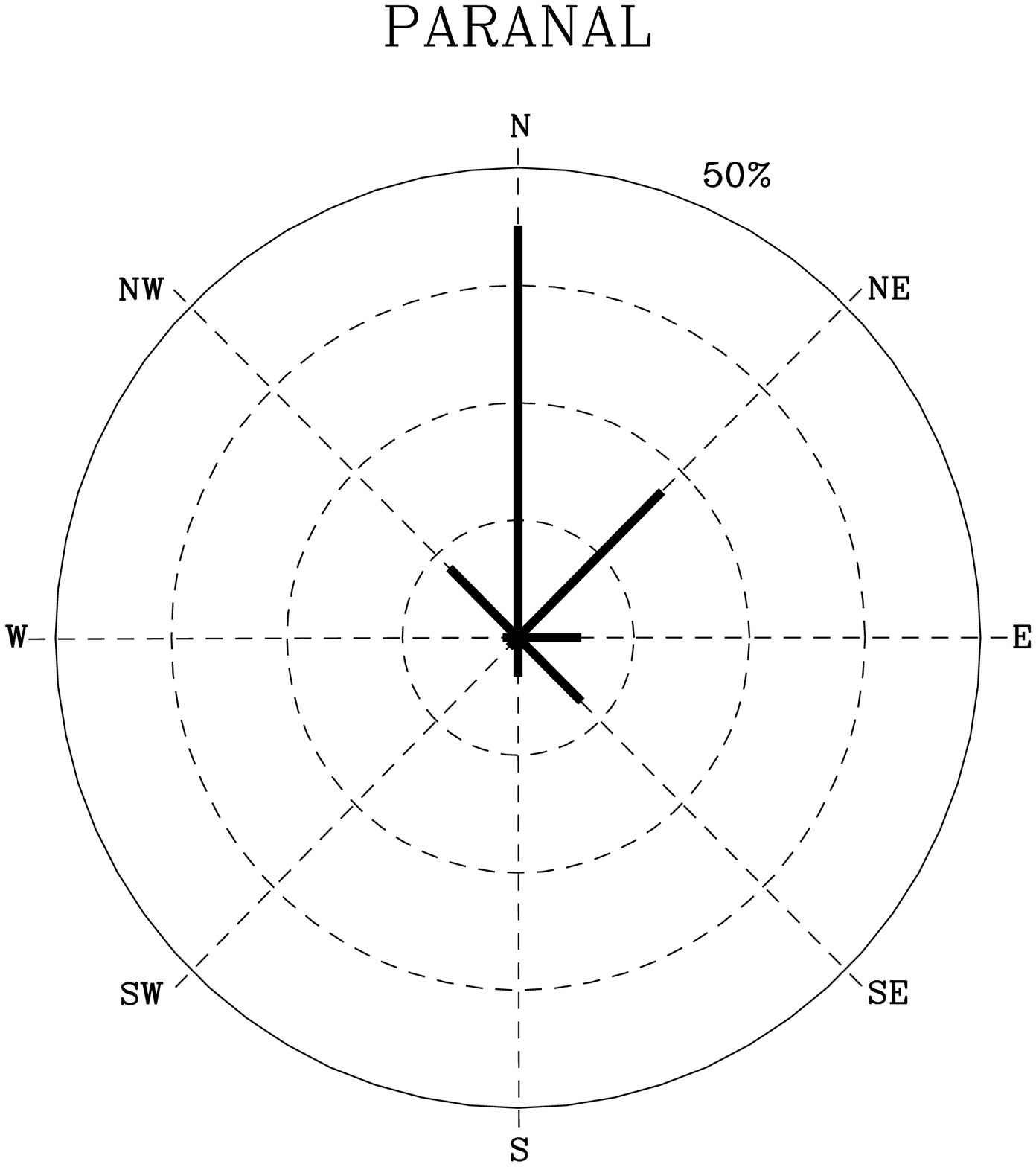}
\includegraphics[width=4cm]{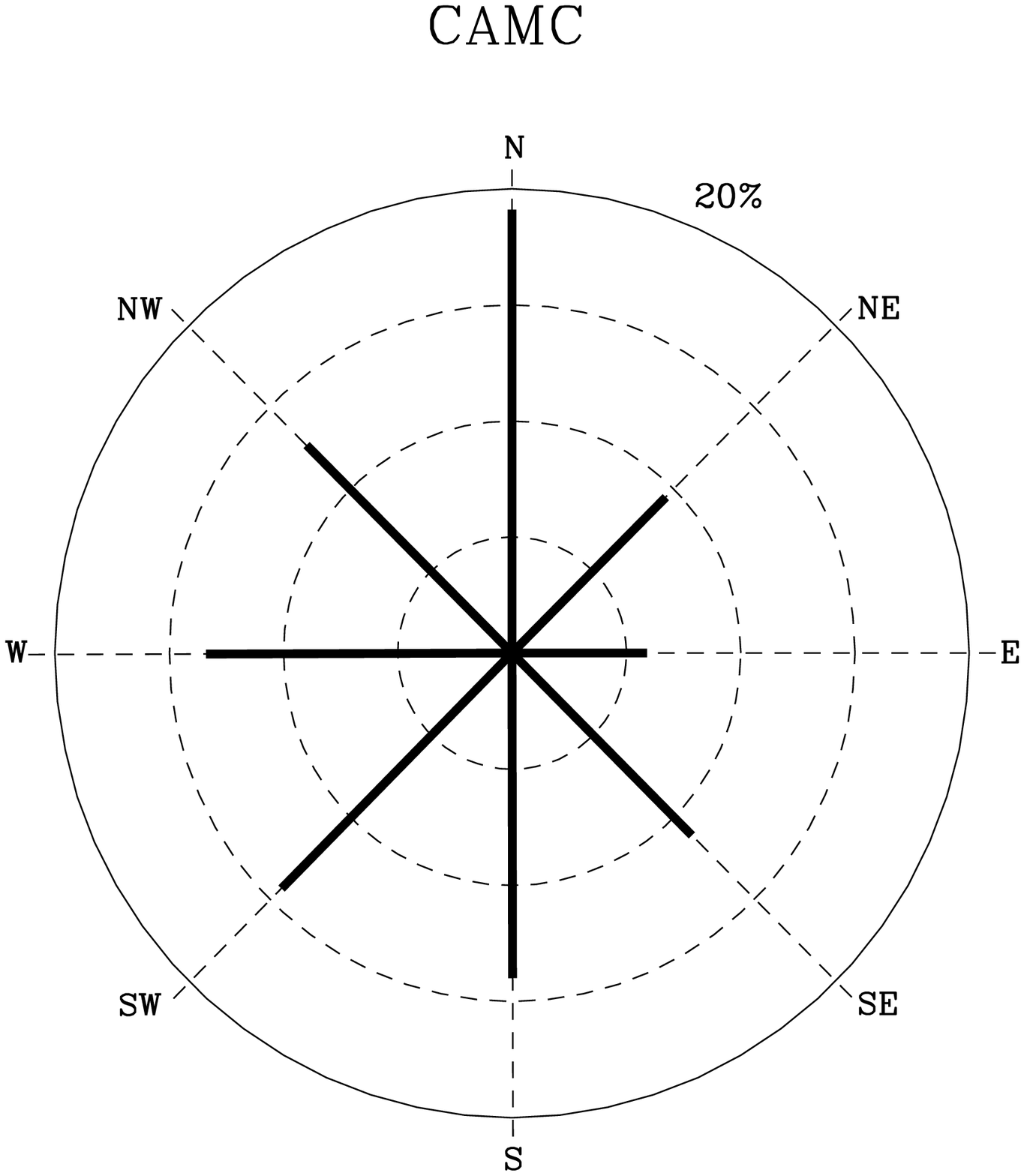}
\includegraphics[width=4cm]{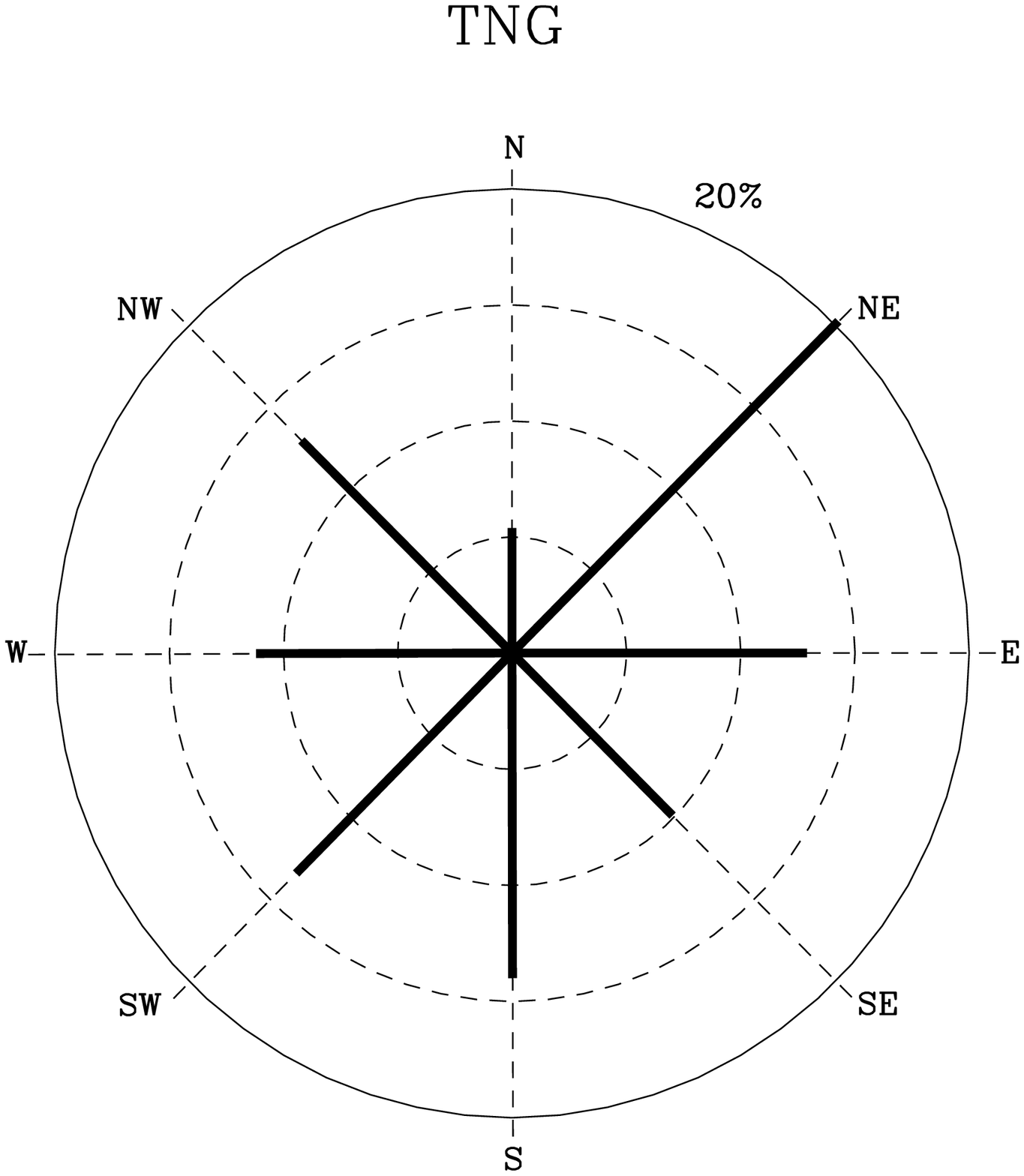}
\includegraphics[width=4cm]{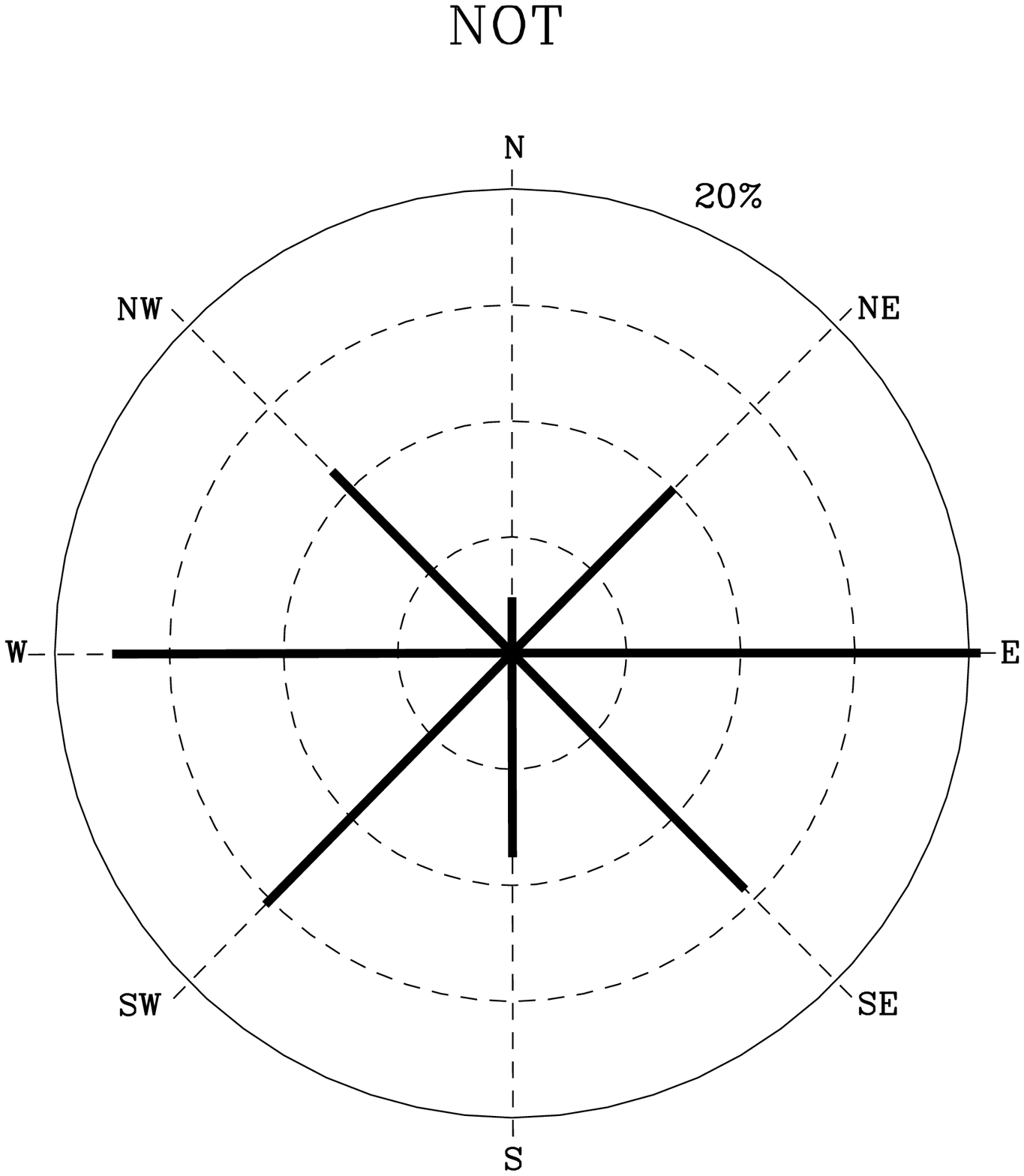}
\caption{Nighttime wind roses at Paranal (1998-2006), and CAMC, TNG
and NOT (1998-2004).} \label{wind roses}
\end{figure}

\begin{figure}%[b]
  \centering
  \includegraphics[width=8cm]{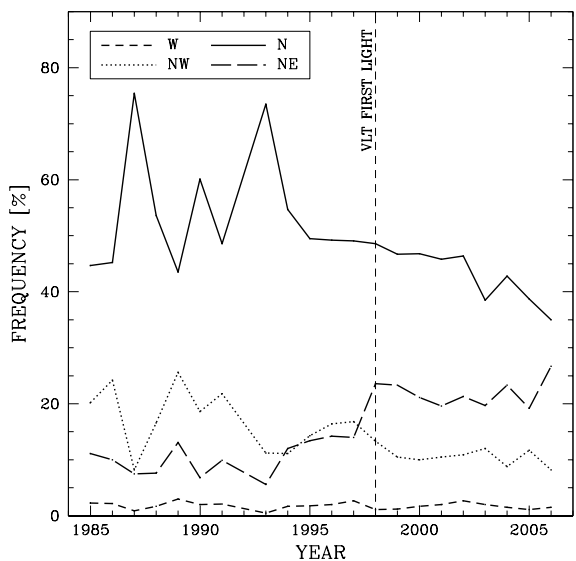}
  \caption{Nighttime evolution of the wind direction through the years at Paranal (quadrant W-NE).}
             \label{winds_N}
   \end{figure}

\begin{figure}%[b]
  \centering
  \includegraphics[width=8cm]{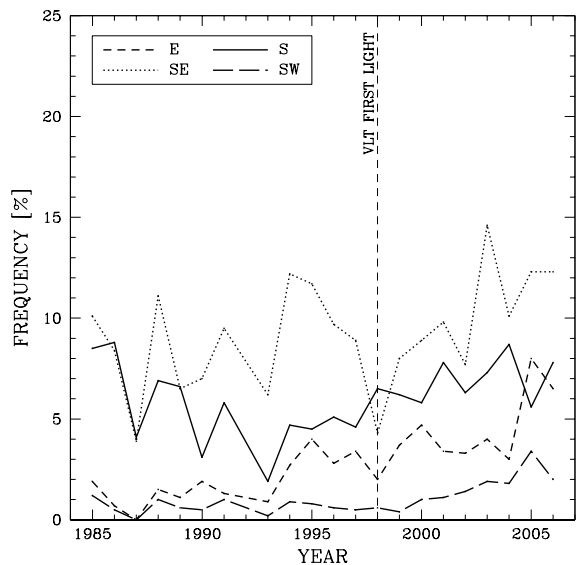}
  \caption{Nighttime evolution of the wind direction through the years at Paranal (quadrant E-SW).}
             \label{winds_S}
   \end{figure}

In the analysis of the wind direction at Paranal we made use of a 9
years database between 1998 (VLT first light) and 2006. In Paper II
the time range used in the computation of the wind statistics at
CAMC was between 1998 and 2004. We consider only nighttime data,
between  22:00 and  4:00 hr Local Time.\\
In Paper II we have demonstrated that there are significant
differences in the wind direction and wind speed ranges among the
ORM. In such paper we considered the locations of the CAMC, the
Telescopio Nazionale Galileo (TNG) and the Nordic Optical Telescope
(NOT). Because of the differences noticed in the three sites, none
of them can be considered as fully representative of the ORM in wind
analysis, so in this case we need to consider them together.
In Figure \ref{wind roses} we show the wind roses computed for Paranal
and CAMC, TNG and NOT (see also Paper II, Figure 2).\\
At CAMC there is no evidence of a prevailing direction. In Paper II
we pointed out that northern winds seem to oscillate with a period of 10 years, while
winds from the north-west show a similar oscillation in the opposite phase.
TNG shows a dominant north-east mode, while NOT has two dominant wind directions (west and east).\\
Mu\~noz-Tu\~n\'on et al. \cite{munoz-tunon98} demonstrated the homogeneity of the image quality
among the ORM. Considering the difference between nighttime wind roses at different locations
in the observatory that we have mentioned above, and considering also the complex orography of the ORM,
should be interesting to investigate if any difference exists in the local Surface Layer seeing
among the observatory. This is an important issue because local conditions have been demonstrated to be crucial
for image quality at Paranal (Lombardi et al.\citealt{lombardi08b} and Sarazin et al.\citealt{sarazin08}).\\
\begin{table}%[t]
    \begin{center}
      \caption[]{Nighttime wind direction frequencies at Paranal from 1998 to 2006.}\scriptsize
   \label{winds}
        \begin{tabular}{ c | c  c  c  c  c  c  c  c }
          \hline
  Year & N & NE & E & SE & S & SW & W & NW \\
           \hline
1998 & 48.6 & 23.6 &  6.5 &  4.3 &  2.0 &  0.6 &  1.1 & 13.3\\
1999 & 46.7 & 23.3 &  6.2 &  8.0 &  3.7 &  0.4 &  1.2 & 10.5\\
2000 & 46.8 & 21.1 &  5.8 &  8.9 &  4.7 &  1.0 &  1.7 & 10.0\\
2001 & 45.8 & 19.6 &  7.8 &  9.8 &  3.4 &  1.1 &  2.0 & 10.5\\
2002 & 46.4 & 21.3 &  6.3 &  7.7 &  3.3 &  1.4 &  2.7 & 10.9\\
2003 & 38.5 & 19.8 &  7.2 & 14.6 &  4.0 &  1.9 &  2.0 & 12.0\\
2004 & 42.8 & 23.3 &  8.7 & 10.1 &  3.0 &  1.8 &  1.5 &  8.8\\
2005 & 38.6 & 19.3 &  5.6 & 12.3 &  8.0 &  3.4 &  1.1 & 11.7\\
2006 & 35.0 & 26.7 &  7.8 & 12.3 &  6.5 &  2.0 &  1.5 &  8.2\\
          \hline
Tot. & 43.8 & 22.0 &  6.8 &  9.6 &  4.2 &  1.4 &  1.7 & 10.5\\
          \hline
        \end{tabular}
        \end{center}
\end{table}
\begin{table}%[t]
    \begin{center}
      \caption[]{Nighttime yearly evolution of the wind direction frequencies at Paranal from 1998 to 2006 as [\% per year].}\scriptsize
   \label{wdir_freq}
        \begin{tabular}{ c  c  c  c  c  c  c  c }
          \hline
   N & NE & E & SE & S & SW & W & NW \\
           \hline
  $-1.6$ & $+0.3$ & $+0.2$ & $+0.9$ & $+0.5$ & $+0.3$ & $<0.1$ & $-0.3$\\
           \hline
        \end{tabular}
        \end{center}
\end{table}
At Paranal a dominant wind blowing from north and north-east appears
during the night. Table \ref{winds} reports the computed wind
direction frequencies at Paranal between 1998 and 2006. The values
in the table are plotted in Figures \ref{winds_N} and \ref{winds_S}
where we include the wind frequencies since 1985 distinguishing two
different epochs before and after the VLT first light occurred in
1998. Wind from north shows a clear decreasing trend through the
years togheter with an increasing of the wind direction from south-east.
This result is in agreement with Sarazin \cite{sarazin04} that in
a previous analysis show a progressively replacement of the
north-westerly wind  by a north-easterly wind. To better investigate this
behavior we have computed the yearly evolution of the nighttime
frequencies of the wind in each direction. There is a strong
oscillation of the wind from north until 1994 (see Figure
\ref{winds_N}). Data from 1998 show a clear trend as shown in Table
\ref{wdir_freq} that reports the computed trend of the yearly evolution
of the frequencies in each
direction. Wind from north is characterized by a decrease of the
frequency of 1.6\% per year marginally compensated by and increasing
of wind coming from south-east (0.9\% per year). We can conclude
that progressively wind coming from the sea is replaced by wind
coming from the Atacama Fault. At the moment it is not clear if we are
in presence of a wide-scale changing of the atmospheric conditions
or if this effect is induced by local conditions.
\subsection{Wind speed}\label{w_sp}
The analysis of the wind speed at Paranal and ORM is carried out
from the calculation of the time in which $w_{sp}$ is in fixed
intervals established on the basis of the safety operations at the
observatories. We consider five main situations:
\begin{itemize}
    \item $w_{sp} < 3$ m s$^{-1}$: negligible wind speed, typically in this case the seeing increases (Paper II);
    \item $3 \leq w_{sp} < 12$ m s$^{-1}$: the wind speed is in the safety range, telescopes observe without
    restriction in the pointing direction, seeing conditions are optimal;
    \item $12 \leq w_{sp} < 15$ m s$^{-1}$: in this interval the telescopes can only point to objects in a direction $\geq 90^{\circ}$ with respect to the actual $w_{dir}$);
    \item $w_{sp} > 15$ m s$^{-1}$: at ORM the telescopes are closed for strong wind;
    \item $w_{sp} > 18$ m s$^{-1}$: at Paranal the telescopes are closed for strong wind.
\end{itemize}
Table \ref{wsp_freq} reports the percentages of time in which the wind speed is in the fixed intervals.
Also in this case, because of the wind speed differences noticed between CAMC, TNG and NOT, none of them can be considered as fully representative of the ORM,
so they have to be taken into account together. For this reason Table \ref{wsp_freq} reports also the results for TNG and NOT obtained in Paper II.\\
\begin{table}%[t]
    \begin{center}
      \caption[]{Nighttime wind speed statistics at Paranal (1998-2006) and CAMC, TNG and NOT (1998-2004).}\scriptsize
   \label{wsp_freq}
        \begin{tabular}{ c  c  c  c  c }
          \hline
   $w_{sp}$ range & Paranal & CAMC & TNG & NOT\\
   $[$m s$^{-1}$ $]$ & $[$\%$]$ & $[$\%$]$ & $[$\%$]$ & $[$\%$]$\\
           \hline
           $w_{sp} < 3$ & 22.1 & 83.6 & 30.2 & 18.5\\
   $3 \leq w_{sp} < 12$ & 68.7 & 16.4 & 68.4 & 70.2\\
  $12 \leq w_{sp} < 15$ &  5.9 &  0.0 &  1.1 &  7.1\\
           \hline
       $w_{sp} \geq 15$ &  3.3 &  0.0 &  0.3 &  4.2\\
       $w_{sp} \geq 18$ & $<0.5$ &  0.0 &  0.0 &  1.2\\
           \hline
        \end{tabular}
        \end{center}
\end{table}
The site of CAMC has a predominance of $w_{sp} < 3$ m s$^{-1}$
($83.6\%$), while Paranal has $22.1\%$, TNG the $30.2\%$ and NOT the
$18.5\%$. Paranal preserves good wind speed conditions in the
$\sim$$70\%$ of the time (same of TNG and NOT), while CAMC only the
$\sim$$16\%$. CAMC never shows $w_{sp} > 12$ m s$^{-1}$, while
Paranal has a percentage of 5.9 in the interval $[12, 15[$ m
s$^{-1}$ and $3.3\%$ of $w_{sp} > 15$ m s$^{-1}$ while not-observing conditions
due to strong wind occur in $< 0.5\%$ of the time. The
TNG has $\sim$$1\%$ of the time with $w_{sp} > 12$ m s$^{-1}$, while
NOT has considerable higher values in the intervals $[12, 15[$
ms$^{-1}$ s$^{-1}$ ($7.1\%$) and $w_{sp} > 15$ m s$^{-1}$ ($4.2\%$).
The highest wind speed measured at Paranal during the years 1998-2006 is
27.4 m s$^{-1}$ in May 2000.

\section{Conclusions}
We have compared the astroclimatological conditions at the Paranal
Observatory and El Roque de Los Muchachos Observatory.
Temperature, air pressure and relative humidity $\sim$$20$ years
trends have been taken into account and significant differences
resulted form the analysis. In particular, CAMC denotes a warming of
$\sim$$1.0^{\circ}$C/10yr, TNG data after 1998 show a flat trend, while the annual temperature trend at
Paranal is characterized by a negative slope between 1993 and 2000
($-0.12$) while data in the period 2000-2006 have a positive slope
(0.19). We have shown that using the pressure height scale both
observatories are dominated by high pressure values and the computed
pressure height scale at Paranal is 8334 m while at ORM it is 8325 m.\\
Typically in cold seasons $RH$ is lower than 60\% at CAMC and $15\%$ at Paranal.
In warm seasons $RH$ is lower than 40\% at CAMC and $20\%$ at Paranal.
A standard requirement for the use of telescopes is a $RH$ value $< 80\%$ or $< 85\%$.
The number of nights in which $RH$ has been higher than the mentioned limits
for more than 50\% of the duration of the night denote a
significant difference between Paranal and CAMC, the first appearing almost immune to high $RH$ events.\\
We have calculated the nighttime annual percentage of time in which
the difference between the air temperature and the dew point
temperature is lower with respect to a variable upper limit $X$ ($X
= 1$ and $5^{\circ}$C).
The analysis has
shown better conditions at Paranal with respect to CAMC in winter,
autumn and spring until 2000, while the two sites are almost equivalent
in the respective summer. Since 2001 the two sites are becoming more similar.\\
We have found a negative correlations between the Southern Oscillation Index
and annual temperatures at Paranal (c.l. 0.7), while no correlation has been found between SOI
and yearly air pressure in the same site.\\
Paranal is typically dominated by wind from north and north-east in
nighttime. The wind speed is lower then 12 m s$^{-1}$ (safety range)
in more than $80\%$ of the time, while it is higher than 18 m
s$^{-1}$ less than the $0.5\%$. The mean wind speed is 7.0 m s$^{-1}$ during the night.
The same analysis at CAMC (see Paper II) shows that the mean wind
speed is 2.2 m s$^{-1}$, but significant differences we found in
both mean direction ad wind speed for each different telescope at ORM.\\

\section*{Acknowledgments}
The authors thank the anonymous reviewer for the very helpful comments.
Gianluca Lombardi also aknowledge the European Southern
Observatory for the availability of the data on-line.

\bsp

\label{lastpage}

\end{document}